# Effect of Electrical Filtering on Level Dependent ASE Noise

**J. Witzens**

Institute of Integrated Photonics, RWTH Aachen University, Sommerfeldstr. 24, 52074 Aachen, Germany

**Abstract:** We derive an analytical model describing the effect of filtering on amplified spontaneous emission noise during or after opto-electronic conversion. In particular, we show that electrical filtering results in a further reduction of the signal quality factor associated with an effective increase of the noise levels and can lead to counter-intuitive dependencies of the measured signal quality on the characteristics of the test setup. Closed form equations are compared with numerical models and experiments, showing excellent agreement.

**Index Terms:** Amplified Spontaneous Emission, Opto-Electronic Receivers

## 1. Introduction

With the advent of the first commercial erbium doped fiber amplifiers (EDFAs) in the late eighties, amplified spontaneous emission (ASE) noise has become an essential aspect of optical communications and is a well-understood phenomenon that has been intensely investigated. Sophisticated models for example take into account deviations from non-Gaussian noise statistics [1, 2], but have been shown to result in very similar signal qualities as predicted by simpler Gaussian models [2, 3]. Here we are taking a closer look at the interaction of signal level dependent ASE noise with receiver (Rx) filtering characteristics and derive a compact set of equations that allow taking these effects into account without complex numerical modeling. These have for example enabled us to model an amplified optical datacom link described in [4] with a comparatively simple model, facilitating system conception and optimization. We show in particular that filtering in the electrical domain reallocates some of the noise power spectral density (PSD) from the 1-level to the 0-level noise and leads to a non-negligible degradation of the optical power budget that needs to be taken into account for accurate system modeling. Furthermore, we show that this effect might lead to subtle discrepancies when characterizing the link with test equipment with different analog bandwidths. For example, discrepancies might arise when recording the signal quality factor (Q-factor) of an optical link with a digital communication analyzer (DCA) and characterizing the link with a bit error rate tester (BERT) if these devices have different analog front-end bandwidths limiting the effective link bandwidth, with these discrepancies reaching beyond expected effects resulting from signal distortion and inter-symbol interference (ISI) or from basic noise filtering. In particular, a reduced bandwidth can, counter-intuitively, result in increased rather than decreased effective noise levels. Predictions are compared to and validated with both numerical models and experiments.

Since ASE noise levels are dependent on instantaneous signal levels, downstream filtering occurring during opto-electronic conversion or in the electric domain inside the Rx can reshape the noise and transfer some noise between the ‘0’ and ‘1’ logical levels leading, as we will show, to an increase of $\sigma_0 + \sigma_1$, the sum of the 0- and 1-level noise standard deviations (std) occurring in the denominator of the signal Q-factor. While the magnitude of this effect can be straightforwardly determined by a numerical simulation of the signal flow, an analytical expression provides superior insight into trends and trade-offs. Here we derive an analytical expression that takes into account both filtering in the optical domain prior to opto-electronic conversion, as well as filtering in the electric domain during or after opto-electronic conversion. To simplify the derivation and for the sake of compactness, filters are assumed to posses an ideal square shaped transfer function, however the derivation can be straightforwardly generalized to an arbitrary filter transfer function. In the next section, a simplified derivation assumes the signal to consist in a single harmonic component oscillating at the Nyquist frequency. The case of a random data stream with a well-defined PSD filtered during or after opto-electronic conversion is treated in the third section. In the fourth part optical filtering is also considered in a comprehensive model. In the fifth section predictions are compared to numerical models and to experiments. Moreover, pattern dependent effects are investigated numerically. Finally, in the sixth and last section the models are generalized to other types of noise occurring in the optical domain prior to opto-electronic conversion such as ASE noise generated prior to optical modulation (the default case treated below consists in optical amplification after modulation) as well as to the case of relative intensity noise (RIN) with a non-uniform PSD rolling off at relatively low frequencies.

## 2. Simplified derivation assuming a single harmonic signal component

Figure 1 depicts the system diagram underpinning the analysis done in the first five sections of this paper. Continuous wave (CW) light is first modulated with a non-return to zero amplitude shift keyed (ASK) signal prior to being optically amplified, optically filtered, transduced into the electric domain and finally electrically filtered.

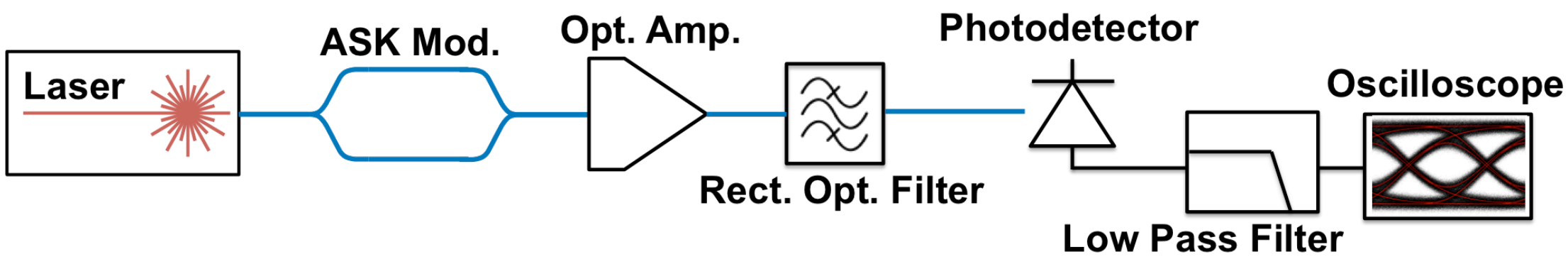

Fig. 1. Diagram of the modeled system: An amplitude modulated signal is optically amplified, fed through a rectangular optical passband filter, transduced, low pass filtered and analyzed for signal quality.

We start by modeling ASE-signal beat noise as white noise modulated by an additional signal. Indeed, for an instantaneous optical power level $P_{SOA}$ entering a semiconductor optical amplifier (SOA) or EDFA, the std of the ASE noise at the output of the SOA, as recorded with an electro-optic Rx with a bandwidth $f_F$, is given by

$$\sigma_{ASE} = \sqrt{2G^2 F h f_0 P_{SOA} f_F} \tag{1}$$

where $G$ is the gain of the SOA, $F$ is its noise factor, $h$ is Planck's constant and $f_0$ is the frequency of the optical carrier. This can be conceptually modeled as white noise with a std given by $\sqrt{2G^2 F h f_0 f_F}$ and a single sided PSD $N_n = 2G^2 F h f_0$ modulated by a time-dependent multiplicative factor $\sqrt{P_{SOA}}$. We denote the Fourier transform of this white noise at a frequency $f_n$ as $\mathcal{F}(f_n)$ (note that the subscript n is introduced here to distinguish pre-filtered noise frequencies from signal frequency components in the more complete analysis reported in section 3).

In order to facilitate the analysis, we assume in this section the time dependency of $\sqrt{P_{SOA}}$ to be described by a single sine wave oscillating at the Nyquist frequency $f_N$, essentially corresponding to a 0101… data pattern. This simplifying assumption is further motivated by two considerations: First, we will show that the higher signal frequency components result in the highest level of transfer between the 0- and 1-level noise and consequently in the worst case estimate of $\sigma_0 + \sigma_1$ (see Eqs. (5) and (7)). Second, the 0101… data pattern also typically results in a signal trace close to the boundaries of the vertical eye opening and thus also constitutes a limiting factor in respect to the bit error rate (BER). Since this data pattern occurs frequently (two sequential bit switches occur ¼ of the time) it is an adequate if somewhat conservative predictor for the BER: a more complete analysis taking a random data stream into account will actually result in a somewhat better estimate of the ASE noise averaged over all 0- or 1-level bit (section 3). A numerical analysis investigating pattern dependent effects will however confirm the coefficients derived here to also hold in the more general case for 010 and 101 patterns.

We parameterize the signal amplitude as

$$\sqrt{P_{SOA}} = \frac{a + b \cdot \cos(2\pi f_N t)}{2} \tag{2}$$

where $a/2$ is the average value of $\sqrt{P_{SOA}}$ and $b = \sqrt{P_{SOA,1}} - \sqrt{P_{SOA,0}}$ is the difference of its 1- and 0-levels and a function of the extinction ratio of the modulator[1].

If we further assume that the electrical filter in the Rx has an ideal square shaped transfer function with a cutoff frequency $f_F$, the total noise after electrical filtering is given by

$$n(t) = \frac{a}{2}\int_{-f_F}^{f_F} \mathcal{F}(f_n)\, e^{i2\pi f_n t} df_n + \frac{b}{4}\int_{-f_F}^{f_F} \mathcal{F}(f_n - f_N)\, e^{i2\pi f_n t} df_n + \frac{b}{4}\int_{-f_F}^{f_F} \mathcal{F}(f_n + f_N)\, e^{i2\pi f_n t} df_n \tag{3}$$

---

[1] Note that while in the case of a Mach-Zehnder modulator (MZM) biased at its quadrature point it might appear more natural to parameterize $P_{SOA}$ instead of $\sqrt{P_{SOA}}$ as a cosine, as the MZM power transfer function has zero second order nonlinearity and $P_{SOA}$ can thus be considered to be approximately proportional to the electrical drive signal (modulo the third order nonlinearity), in the case of a critically coupled resonant ring modulator operated close to resonance as utilized in [4] it is rather $\sqrt{P_{SOA}}$ that is approximately proportional to the drive signal, as the electrical signal to optical power transfer function can be approximated as being a square function in the small signal limit.

$$= \frac{a}{2}\int_{-f_F}^{f_F} \mathcal{F}(f_n)\, e^{i2\pi f_n t} df_n + \frac{b}{4}\int_{-f_F-f_N}^{f_F-f_N} \mathcal{F}(f_n)\, e^{i2\pi (f_n+f_N)t} df_n + \frac{b}{4}\int_{-f_F+f_N}^{f_F+f_N} \mathcal{F}(f_n)\, e^{i2\pi (f_n-f_N)t} df_n$$

where $t$ is the time at which the filtered noise is calculated. Equation (3) describes the up- and down-conversion of the noise by multiplication with the signal amplitude followed by electrical filtering. Evaluated at $t = 0$, i.e., at a time where $\sqrt{P_{SOA}} = (a+b)/2$ corresponds to the 1-level, and assuming $f_F > f_N$ (which is typically the case in a functional Rx), this integral results in

$$n(t) = \frac{a+b}{2}\int_{-f_F+f_N}^{f_F-f_N} \mathcal{F}(f_n)\, df_n + \frac{2a+b}{4}\int_{-f_F}^{-f_F+f_N} \mathcal{F}(f_n)\, df_n + \frac{2a+b}{4}\int_{f_F-f_N}^{f_F} \mathcal{F}(f_n)\, df_n + \frac{b}{4}\int_{-f_F-f_N}^{-f_F} \mathcal{F}(f_n)\, df_n + \frac{b}{4}\int_{f_F}^{f_F+f_N} \mathcal{F}(f_n)\, df_n \tag{4}$$

After integration of the PSD, this results in a std given by

$$\begin{aligned}
\sigma_1 &= \sqrt{N_n \cdot \left[\left(\frac{a+b}{2}\right)^2 (f_F - f_N) + \left(\frac{2a+b}{4}\right)^2 f_N + \left(\frac{b}{4}\right)^2 f_N\right]} \\
&= \sqrt{N_n \cdot \left[P_{SOA,1}(f_F - f_N) + \left(\sqrt{P_{SOA,1}} - \frac{\sqrt{P_{SOA,1}}-\sqrt{P_{SOA,0}}}{4}\right)^2 f_N + \left(\frac{\sqrt{P_{SOA,1}}-\sqrt{P_{SOA,0}}}{4}\right)^2 f_N\right]} \\
&= \sqrt{N_n \cdot \left[P_{SOA,1} f_F - \tfrac{1}{2}\sqrt{P_{SOA,1}}\left(\sqrt{P_{SOA,1}} - \sqrt{P_{SOA,0}}\right) f_N + \tfrac{1}{8}\left(\sqrt{P_{SOA,1}} - \sqrt{P_{SOA,0}}\right)^2 f_N\right]} \\
&= \sqrt{N_n \cdot \left[\begin{matrix} P_{SOA,1} f_F - \tfrac{1}{4}\left(\left(\sqrt{P_{SOA,1}} + \sqrt{P_{SOA,0}}\right) + \left(\sqrt{P_{SOA,1}} - \sqrt{P_{SOA,0}}\right)\right)\left(\sqrt{P_{SOA,1}} - \sqrt{P_{SOA,0}}\right) f_N \\ + \tfrac{1}{8}\left(\sqrt{P_{SOA,1}} - \sqrt{P_{SOA,0}}\right)^2 f_N \end{matrix}\right]} \\
&= \sqrt{2G^2 F h f_0 \cdot \left[P_{SOA,1} f_F - \tfrac{1}{4}\left(P_{SOA,1} - P_{SOA,0}\right) f_N - \tfrac{1}{8}\left(\sqrt{P_{SOA,1}} - \sqrt{P_{SOA,0}}\right)^2 f_N\right]}
\end{aligned} \tag{5}$$

It can be straightforwardly verified that this formula reduces to the usual formula for ASE-signal beat noise $\sqrt{2G^2 F h f_0 P_{SOA,1} f_F}$ for either $f_N \to 0$ (slowly varying signal) or for $P_{SOA,0} = P_{SOA,1}$ (constant signal level).

Conversely, when Eq. (3) is evaluated at $t = 1/2f_N$, i.e., at a time where $\sqrt{P_{SOA}} = (a-b)/2$ corresponds to a logical zero, and assuming as previously $f_F > f_N$, it reduces to

$$n(t) = \frac{a-b}{2}\int_{-f_F+f_N}^{f_F-f_N} \mathcal{F}(f_n)\, df_n + \frac{2a-b}{4}\int_{-f_F}^{-f_F+f_N} \mathcal{F}(f_n)\, df_n + \frac{2a-b}{4}\int_{f_F-f_N}^{f_F} \mathcal{F}(f_n)\, df_n - \frac{b}{4}\int_{-f_F-f_N}^{-f_F} \mathcal{F}(f_n)\, df_n - \frac{b}{4}\int_{f_F}^{f_F+f_N} \mathcal{F}(f_n)\, df_n \tag{6}$$

Following a similar derivation as for (5), this results in a noise std given by

$$\begin{aligned}
\sigma_0 &= \sqrt{N_n \cdot \left[\left(\frac{a-b}{2}\right)^2 (f_F - f_N) + \left(\frac{2a-b}{4}\right)^2 f_N + \left(\frac{b}{4}\right)^2 f_N\right]} \\
&= \sqrt{2G^2 F h f_0 \cdot \left[P_{SOA,0} f_F + \frac{1}{4}\left(P_{SOA,1} - P_{SOA,0}\right) f_N - \frac{1}{8}\left(\sqrt{P_{SOA,1}} - \sqrt{P_{SOA,0}}\right)^2 f_N\right]}
\end{aligned} \tag{7}$$

Here too the std reduces to the usual formula when either $f_N \to 0$ or $P_{SOA,0} = P_{SOA,1}$.

Compared to the ASE noise levels occurring for a slowly varying signal ($f_N \to 0$) filtered with an electrical filter with a noise equivalent bandwidth $f_F$, $\overline{\sigma_{0/1}} = \sqrt{2G^2 F h f_0 P_{SOA,0/1} f_F}$, there are two differences: First, there is a net reduction of both the noise variances corresponding to the term $-\frac{1}{8}\left(\sqrt{P_{SOA,1}} - \sqrt{P_{SOA,0}}\right)^2 f_N$ in Eqs. (5) and (7). Second, there is a transfer of noise between the variance of the 1-level noise (that is reduced) and the variance of the 0-level noise (that is increased).

Importantly, the net effect is to increase the effective amount of noise and to reduce the overall signal Q-factor as compared to a less sophisticated model with noise estimates based solely on $\overline{\sigma_0}$ and $\overline{\sigma_1}$:

In order to show this, we first recast Eqs. (5) and (7) into

$$\sigma_{0/1} = \sqrt{2G^2Fhf_0 \cdot \left[P_{SOA,0/1}f_F \pm 0.5\sqrt{P_{SOA,0/1}}\left(\sqrt{P_{SOA,1}} - \sqrt{P_{SOA,0}}\right)f_N + \frac{\left(\sqrt{P_{SOA,1}}-\sqrt{P_{SOA,0}}\right)^2}{8}f_N\right]} \tag{8}$$

(this actually corresponds to the third line in the derivation of Eq. (5)). In order to show that $\sigma_0 + \sigma_1$ is larger than $\overline{\sigma_0} + \overline{\sigma_1}$ we develop $\overline{\sigma_0}$ and $\overline{\sigma_1}$ as first order Taylor series and use the fact that the square root function is downward concave to obtain inequalities yielding an upper bound for $\overline{\sigma_0} + \overline{\sigma_1}$, i.e.,

$$\overline{\sigma_0} = \sqrt{\sigma_0^2 - N_n \cdot \left[0.5\sqrt{P_{SOA,0}}\left(\sqrt{P_{SOA,1}} - \sqrt{P_{SOA,0}}\right)f_N + \frac{\left(\sqrt{P_{SOA,1}}-\sqrt{P_{SOA,0}}\right)^2}{8}f_N\right]} \tag{9}$$

results in

$$\overline{\sigma_0} < \sigma_0 - \frac{1}{4}\frac{\sqrt{N_n}\cdot\left(\sqrt{P_{SOA,1}}-\sqrt{P_{SOA,0}}\right)f_N}{\sqrt{f_F}} - \frac{1}{16}\frac{N_n\left(\sqrt{P_{SOA,1}}-\sqrt{P_{SOA,0}}\right)^2 f_N}{\sigma_0} \tag{10}$$

Similarly

$$\overline{\sigma_1} < \sigma_1 + \frac{1}{4}\frac{\sqrt{N_n}\cdot\left(\sqrt{P_{SOA,1}}-\sqrt{P_{SOA,0}}\right)f_N}{\sqrt{f_F}} - \frac{1}{16}\frac{N_n\left(\sqrt{P_{SOA,1}}-\sqrt{P_{SOA,0}}\right)^2 f_N}{\sigma_1} \tag{11}$$

and thus

$$\overline{\sigma_0} + \overline{\sigma_1} < \sigma_0 + \sigma_1 - \frac{1}{16}N_n\left(\sqrt{P_{SOA,1}} - \sqrt{P_{SOA,0}}\right)^2 f_N\left(\frac{1}{\sigma_0} + \frac{1}{\sigma_1}\right) \tag{12}$$

In other words, $\sigma_0 + \sigma_1$ is clearly larger than $\overline{\sigma_0} + \overline{\sigma_1}$ so that the net effect of the modification of level dependent noise by electrical filtering inside the Rx (further referred to as "noise mixing") is to further decrease the signal Q-factor.

## 3. Derivation assuming a random data stream

We now move to a more general model taking the complete signal spectrum of a random amplitude modulated data stream into account. Following a similar notation as in the previous section, the signal is expressed as a sum of its Fourier components as

$$\sqrt{P_{SOA}} = \frac{a}{2} + \int_{-\infty}^{\infty}\frac{b(f_s)}{4}e^{i2\pi f_s t}df_s \tag{13}$$

where $f_s$ denotes the signal frequencies and $b(-f_s) = b(f_s)^*$ since the signal is real valued. Returning to Eqs. (4) and (6), the std of the 0- and 1-level ASE-signal beat noise are expressed as

$$\sigma_{0/1} = \sqrt{N_n \cdot \left[\int_{f_n=0}^{f_n=f_F}\left|\frac{2a \mp \int_{f_s=0}^{f_s=f_F-f_n}2Re(b)\,df_s \mp \int_{f_s=f_F-f_n}^{f_s=f_F+f_n}b\,df_s}{4}\right|^2 df_n + \int_{f_n=f_F}^{f_n=f_{max}+f_F}\left|\frac{\int_{f_s=f_n-f_F}^{f_s=f_n+f_F}b\,df_s}{4}\right|^2 df_n\right]} \tag{14}$$

where $f_n$ is the white noise frequency before up or down-conversion by multiplication with $\sqrt{P_{SOA}}$. We further note the maximum frequency at which the signal has a non-zero PSD as $f_{max}$. Without loss of generality we can assume that $t = 0$ corresponds to the sampling time of a 1-level bit, in which case $\frac{a}{2} + \int_0^{f_{max}}2\frac{Re(b)}{4}df_S = \sqrt{P_{SOA,1}}$ (evaluation of Eq. (13) at $t = 0$) and $\frac{a}{2} - \int_0^{f_{max}}2\frac{Re(b)}{4}df_S = 2 \cdot mean\left(\sqrt{P_{SOA}}\right) - \sqrt{P_{SOA,1}} = \sqrt{P_{SOA,0}}$ (the signal is assumed to be DC balanced). Using these relations and applying a change of variables given by $f_n' = f_n - f_F$ to the second term we obtain

$$\sigma_{0/1} = \sqrt{N_n \cdot \left[\int_{f_n=0}^{f_n=f_F} \left|\frac{4\sqrt{P_{SOA,0/1}} \pm \int_{f_s=f_F-f_n}^{f_s=f_{max}} 2Re(b)\, df_s \mp \int_{f_s=f_F-f_n}^{f_s=f_F+f_n} b df_s}{4}\right|^2 df_n + \int_{f_n=0}^{f_n=f_{max}} \left|\frac{\int_{f_s=f_n}^{f_s=f_n+2f_F} b\, df_s}{4}\right|^2 df_n\right]} \tag{15}$$

Applying another change of variables given by $f_n' = f_F - f_n$ to the first term we further transform this expression into

$$\sigma_{0/1} = \sqrt{N_n \cdot \left[\int_{f_n=0}^{f_n=f_F} \left|\frac{4\sqrt{P_{SOA,0/1}} \pm \int_{f_s=f_n}^{f_s=f_{max}} 2Re(b)\, df_s \mp \int_{f_s=f_n}^{f_s=2f_F-f_n} b df_s}{4}\right|^2 df_n + \int_{f_n=0}^{f_n=f_{max}} \left|\frac{\int_{f_s=f_n}^{f_s=f_n+2f_F} b\, df_s}{4}\right|^2 df_n\right]} \tag{16}$$

Prior to a more general derivation (Eq. (29) and below), as a first step we make the simplifying assumption $f_F \geq f_{max}$ (i.e., the electrical filter has no effect on the signal). This assumption allows us to replace both of the upper integration bounds $2f_F - f_n$ and $f_n + 2f_F$ by $f_{max}$ and to simplify (16) as

$$\sigma_{0/1} = \sqrt{N_n \cdot \left[\int_{f_n=0}^{f_n=f_F} \left|\frac{4\sqrt{P_{SOA,0/1}} \pm \int_{f_s=f_n}^{f_s=f_{max}} b^*\, df_s}{4}\right|^2 df_n + \int_{f_n=0}^{f_n=f_{max}} \left|\frac{\int_{f_s=f_n}^{f_s=f_{max}} b\, df_s}{4}\right|^2 df_n\right]} \tag{17}$$

which is further developed into

$$\begin{aligned} {\sigma_{0/1}}^2/N_n &= P_{SOA,0/1} f_F \pm \frac{1}{2}\sqrt{P_{SOA,0/1}} \int_{f_s=0}^{f_s=f_{max}} Re(b) f_s\, df_s \\ &+ \frac{1}{8}\left(\int_{f_{s1}=0}^{f_{s1}=f_{max}} \int_{f_{s2}=0}^{f_{s2}=f_{max}} min(f_{s1}, f_{s2}) \left(Re\big(b(f_{s1})\big) Re\big(b(f_{s2})\big) + Im\big(b(f_{s1})\big) Im\big(b(f_{s2})\big)\right) df_{s1} df_{s2}\right) \end{aligned} \tag{18}$$

A further treatment requires the evaluation of the integral

$$\frac{\int_{f_s=0}^{f_s=f_{max}} Re(b) f_s\, df_s}{\int_{f_s=0}^{f_s=f_{max}} Re(b)\, df_s} = \frac{\int_{f_s=0}^{f_s=f_{max}} Re(b) f_s\, df_s}{\sqrt{P_{SOA,1}} - \sqrt{P_{SOA,0}}} = \mu f_N \tag{19}$$

which requires an assumption on the specific signal shape (the last term simply corresponds to the definition of $\mu$). For the numerical evaluation of this integral we assume the random data stream to have a single sided PSD given by

$$N_s = \left(\frac{\sqrt{P_{SOA,1}} + \sqrt{P_{SOA,0}}}{2}\right)^2 \delta(f_s = 0) + \left(\frac{\sqrt{P_{SOA,1}} - \sqrt{P_{SOA,0}}}{2}\right)^2 \frac{2}{DK} \frac{sin(\pi f_s/D)^2}{(\pi f_s/D)^2} \tag{20}$$

for signal frequencies verifying $f_s \leq f_{max}$ and equal to zero (filtered) for frequencies above $f_{max}$, where $D = 2f_N$ is the data rate. $K$ is a normalization factor taking into account the truncation of the distribution (for completeness as it is of no further relevance in the following derivations). Under these conditions we obtain an expectation value for $\mu$ given by

$$\langle\mu\rangle = \frac{\int_0^{\pi\frac{f_{max}}{2f_N}} \frac{sin(x)}{x} \frac{2x}{\pi} dx}{\int_0^{\pi\frac{f_{max}}{2f_N}} \frac{sin(x)}{x} dx} \cong 0.69 \tag{21}$$

where the numerical estimate was done for $f_{max} = 2f_N$.

The exact value of $\mu$ depends not only on the global bit sequence, but also on the position of the specific bit at which it is evaluated within the global bit sequence (i.e., which bit is chosen to correspond to $t = 0$) pointing to the fact that the magnitude of the noise mixing depends on the local data pattern (this is further discussed in section 5). Nonetheless, its average effect on the 0- and 1-level noise can be evaluated with the expectation value of $\mu$. Since $\mu$ also depends on the exact shape of the bits, it needs to be adapted to the specific signal PSD. The numerical estimate in Eq. (21) is thus to be only understood as a typical value. A systematic discussion follows:

Interestingly the noise mixing term (i.e., as described by $\mu$ in Eq. (28)) can be completely cancelled for certain data patterns: For example, if the distribution given by Eq. (20) is filtered for frequencies above $4f_N$ instead of the $2f_N$ cutoff assumed in Eq. (21), the expectation value of $\mu$ would be zero (however, in practice the latter case corresponds to a more typical situation). Figure 2 shows the numerical evaluation of $\langle\mu\rangle$ for the signal PSD given by Eq. (20) as a function of $f_{max}$ expressed as a multiple of $f_N$.

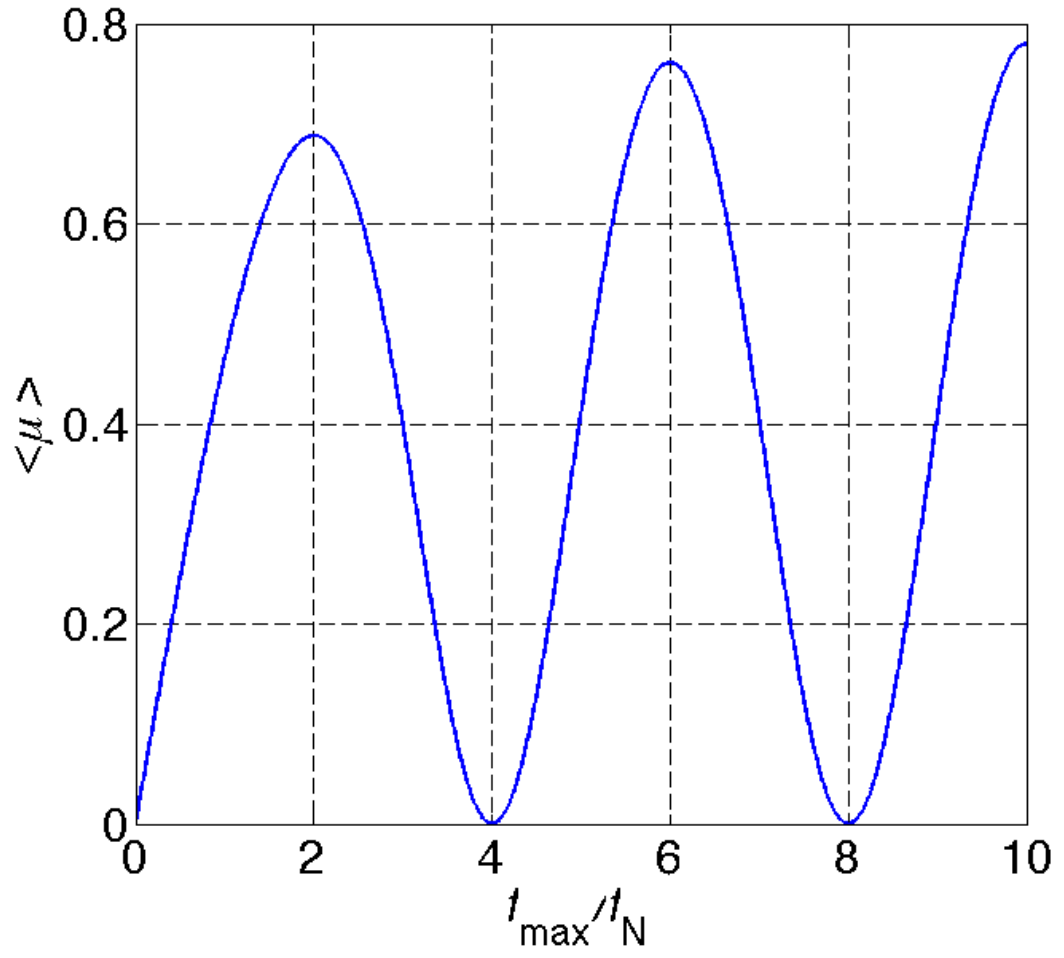

Fig. 2. Numerical evaluation of the noise mixing parameter $\mu$ as a function of $f_{max}/f_N$ assuming the signal PSD given by Eq. (20).

Furthermore, we also need to evaluate the following integral also occurring in Eq. (18)

$$\int_{f_{s1}=0}^{f_{s1}=f_{max}}\int_{f_{s2}=0}^{f_{s2}=f_{max}} min(f_{s1},f_{s2})\left(Re\big(b(f_{s1})\big)Re\big(b(f_{s2})\big)+Im\big(b(f_{s1})\big)Im\big(b(f_{s2})\big)\right)df_{s1}df_{s2} = \left(\sqrt{P_{SOA,1}}-\sqrt{P_{SOA,0}}\right)^2(2\mu-\gamma)f_N \tag{22}$$

where the equation introduces the newly defined coefficient $\gamma$. This integral can be further expressed as the sum of two other integrals as follows

$$\begin{aligned}&\int_{f_{s1}=0}^{f_{s1}=f_{max}}\int_{f_{s2}=0}^{f_{s2}=f_{max}}\frac{f_{s1}+f_{s2}}{2}\left(Re\big(b(f_{s1})\big)Re\big(b(f_{s2})\big)+Im\big(b(f_{s1})\big)Im\big(b(f_{s2})\big)\right)df_{s1}df_{s2}\\&-\int_{f_{s1}=0}^{f_{s1}=f_{max}}\int_{f_{s2}=0}^{f_{s2}=f_{max}}\left|\frac{f_{s1}-f_{s2}}{2}\right|\left(Re\big(b(f_{s1})\big)Re\big(b(f_{s2})\big)+Im\big(b(f_{s1})\big)Im\big(b(f_{s2})\big)\right)df_{s1}df_{s2}\\&=\int_{f_{s1}=0}^{f_{s1}=f_{max}}\int_{f_{s2}=0}^{f_{s2}=f_{max}}f_{s1}\left(Re\big(b(f_{s1})\big)Re\big(b(f_{s2})\big)+Im\big(b(f_{s1})\big)Im\big(b(f_{s2})\big)\right)df_{s1}df_{s2}\\&-\int_{f_{s1}=0}^{f_{s1}=f_{max}}\int_{f_{s2}=0}^{f_{s2}=f_{max}}\left|\frac{f_{s1}-f_{s2}}{2}\right|\left(Re\big(b(f_{s1})\big)Re\big(b(f_{s2})\big)+Im\big(b(f_{s1})\big)Im\big(b(f_{s2})\big)\right)df_{s1}df_{s2}\end{aligned} \tag{23}$$

These two terms are treated separately. The first integral is straightforward to evaluate as

$$\begin{aligned}&\int_{f_{s1}=0}^{f_{s1}=f_{max}}\int_{f_{s2}=0}^{f_{s2}=f_{max}}f_{s1}\left(Re\big(b(f_{s1})\big)Re\big(b(f_{s2})\big)+Im\big(b(f_{s1})\big)Im\big(b(f_{s2})\big)\right)df_{s1}df_{s2}\\&=\int_{f_{s1}=0}^{f_{s1}=f_{max}}f_{s1}Re\big(b(f_{s1})\big)df_{s1}\int_{f_{s2}=0}^{f_{s2}=f_{max}}Re\big(b(f_{s2})\big)df_{s2}+\int_{f_{s1}=0}^{f_{s1}=f_{max}}f_{s1}Im\big(b(f_{s1})\big)df_{s1}\int_{f_{s2}=0}^{f_{s2}=f_{max}}Im\big(b(f_{s2})\big)df_{s2}\\&=\left(\sqrt{P_{SOA,1}}-\sqrt{P_{SOA,0}}\right)^2\mu f_N\end{aligned} \tag{24}$$

The term $\int_{f_{s2}=0}^{f_{s2}=f_{max}} Im\big(b(f_{s2})\big)df_{s2}$ is zero since it is proportional to the derivative of the signal at the sampling time (assuming sampling to occur when the 1-level is maximized, respectively when the 0-level is minimized, or when they reach a plateau). Starting from Eq. (13) the derivative of the signal is given by

$$\frac{d\sqrt{P_{SOA}}}{dt} = \int_0^{f_{max}} \left(-\frac{Re\big(b(f_s)\big)}{2}sin(2\pi f_s t) - \frac{Im\big(b(f_s)\big)}{2}cos(2\pi f_s t)\right)2\pi df_s \tag{25}$$

Evaluation of this equation at $t = 0$ yields $\int_{f_{s2}=0}^{f_{s2}=f_{max}} Im\big(b(f_{s2})\big)df_{s2} = 0$ assuming $\left(\frac{d\sqrt{P_{SOA}}}{dt}\right)_{t=0} = 0$.

The second integral from Eq. (23) is somewhat more complex to handle and is treated as follows

$$\begin{aligned}
&\int_{f_{s1}=0}^{f_{s1}=f_{max}}\int_{f_{s2}=0}^{f_{s2}=f_{max}} \left|\frac{f_{s1}-f_{s2}}{2}\right| \Big(Re\big(b(f_{s1})\big)Re\big(b(f_{s2})\big) + Im\big(b(f_{s1})\big)Im\big(b(f_{s2})\big)\Big)df_{s1}df_{s2} \\
&= \int_{f_{s1}=-f_{max}}^{f_{s1}=f_{max}}\int_{f_{s2}=-f_{max}}^{f_{s2}=f_{max}} \left|\frac{f_{s1}-f_{s2}}{4}\right| \Big(Re\big(b(f_{s1})\big)Re\big(b(f_{s2})\big) + Im\big(b(f_{s1})\big)Im\big(b(f_{s2})\big)\Big)df_{s1}df_{s2} \\
&\quad - \int_{f_{s1}=-f_{max}}^{f_{s1}=0}\int_{f_{s2}=0}^{f_{s2}=f_{max}} \left|\frac{f_{s1}-f_{s2}}{4}\right| \Big(Re\big(b(f_{s1})\big)Re\big(b(f_{s2})\big) + Im\big(b(f_{s1})\big)Im\big(b(f_{s2})\big)\Big)df_{s1}df_{s2} \\
&\quad - \int_{f_{s1}=0}^{f_{s1}=f_{max}}\int_{f_{s2}=-f_{max}}^{f_{s2}=0} \left|\frac{f_{s1}-f_{s2}}{4}\right| \Big(Re\big(b(f_{s1})\big)Re\big(b(f_{s2})\big) + Im\big(b(f_{s1})\big)Im\big(b(f_{s2})\big)\Big)df_{s1}df_{s2} \\
&= \int_{f_{s1}=-f_{max}}^{f_{s1}=f_{max}}\int_{f_{s2}=-f_{max}}^{f_{s2}=f_{max}} \left|\frac{f_{s1}-f_{s2}}{4}\right| \Big(Re\big(b(f_{s1})\big)Re\big(b(f_{s2})\big) + Im\big(b(f_{s1})\big)Im\big(b(f_{s2})\big)\Big)df_{s1}df_{s2} \\
&\quad - \int_{f_{s1}=0}^{f_{s1}=f_{max}}\int_{f_{s2}=0}^{f_{s2}=f_{max}} \left(\frac{f_{s1}+f_{s2}}{2}\right) \Big(Re\big(b(f_{s1})\big)Re\big(b(f_{s2})\big) - Im\big(b(f_{s1})\big)Im\big(b(f_{s2})\big)\Big)df_{s1}df_{s2} \\
&= \int_{f_{s1}=-f_{max}}^{f_{s1}=f_{max}}\int_{f_{s2}=-f_{max}}^{f_{s2}=f_{max}} \left|\frac{f_{s1}-f_{s2}}{4}\right| \Big(Re\big(b(f_{s1})\big)Re\big(b(f_{s2})\big) + Im\big(b(f_{s1})\big)Im\big(b(f_{s2})\big)\Big)df_{s1}df_{s2} \\
&\quad -\left(\sqrt{P_{SOA,1}} - \sqrt{P_{SOA,0}}\right)^2 \mu f_N
\end{aligned} \tag{26}$$

In the second equality we used $Re\big(b(-f_s)\big) = Re\big(b(f_s)\big)$ and $Im\big(b(-f_s)\big) = -Im\big(b(f_s)\big)$.

We now apply a change of variable $f = f_{s1} - f_{s2}$ to the remaining integral that then takes the form of a convolution of two Fourier transforms and can thus be expressed as the Fourier transform of a product of two functions

$$\begin{aligned}
&\left(\sqrt{P_{SOA,1}} - \sqrt{P_{SOA,0}}\right)^2 \gamma f_N \\
&= \int_{f_{s1}=-f_{max}}^{f_{s1}=f_{max}}\int_{f_{s2}=-f_{max}}^{f_{s2}=f_{max}} \left|\frac{f_{s1}-f_{s2}}{4}\right| \Big(Re\big(b(f_{s1})\big)Re\big(b(f_{s2})\big) + Im\big(b(f_{s1})\big)Im\big(b(f_{s2})\big)\Big)df_{s1}df_{s2} \\
&= \int_{f=-\infty}^{f=\infty}\int_{f_{s1}=-\infty}^{f_{s1}=\infty} 4|f| \left(\frac{Re\big(b(f_{s1})\big)}{4}\frac{Re\big(b(f-f_{s1})\big)}{4} - \frac{Im\big(b(f_{s1})\big)}{4}\frac{Im\big(b(f-f_{s1})\big)}{4}\right)df_{s1}df \\
&= 8\int_{f=0}^{f=2f_{max}} |f| Re\left(\mathcal{F}\left(\left(\sqrt{P_{SOA}} - mean\left(\sqrt{P_{SOA}}\right)\right)^2\right)\right)df
\end{aligned} \tag{27}$$

With $\mathcal{F}$ denoting here the Fourier operator (one may straightforwardly verify that this reduces to $\gamma = 1$ under the assumption of the simple signal shape given by Eq. (2)).

Returning to Eq. (18) we can now rewrite the 0- and 1-level noise std as

$$\sigma_{0/1}{}^2/N_n = P_{SOA,0/1}f_F \pm \frac{1}{2}\sqrt{P_{SOA,0/1}}\left(\sqrt{P_{SOA,1}} - \sqrt{P_{SOA,0}}\right)\mu f_N + \frac{1}{8}\left(\sqrt{P_{SOA,1}} - \sqrt{P_{SOA,0}}\right)^2 (2\mu - \gamma)f_N$$
$$= P_{SOA,0/1}f_F \pm \frac{1}{4}\left(P_{SOA,1} - P_{SOA,0}\right)\mu f_N - \frac{1}{8}\left(\sqrt{P_{SOA,1}} - \sqrt{P_{SOA,0}}\right)^2 \gamma f_N \quad (28)$$

Eq. (28) is similar to Eqs. (5) and (7) derived in the simpler context of a signal with a single Fourier component at a frequency $f_s = f_N$, with the differences that $f_N$ is replaced by $\mu f_N$ in the second term and by $\gamma f_N$ in the third term. In that sense $\mu f_N$ and $\gamma f_N$ correspond to weighted averages of the signal frequencies.

Equation (27) is numerically evaluated assuming the signal PSD given by Eq. (20) for different values of $f_{max}/f_N$. For each value of $f_{max}/f_N$ the expectation value $\langle\gamma\rangle$ is evaluated based on 10000 randomly generated data streams of 300 bits each. It can be seen in Fig. 3 that like $\langle\mu\rangle$, $\langle\gamma\rangle$ features an oscillatory behavior as a function of $f_{max}/f_N$. Moreover, while at certain values such as $f_{max} = 2f_N$ and $f_{max} = 4f_N$ $\langle\mu\rangle$ and $\langle\gamma\rangle$ are close to each other, in general they significantly differ.

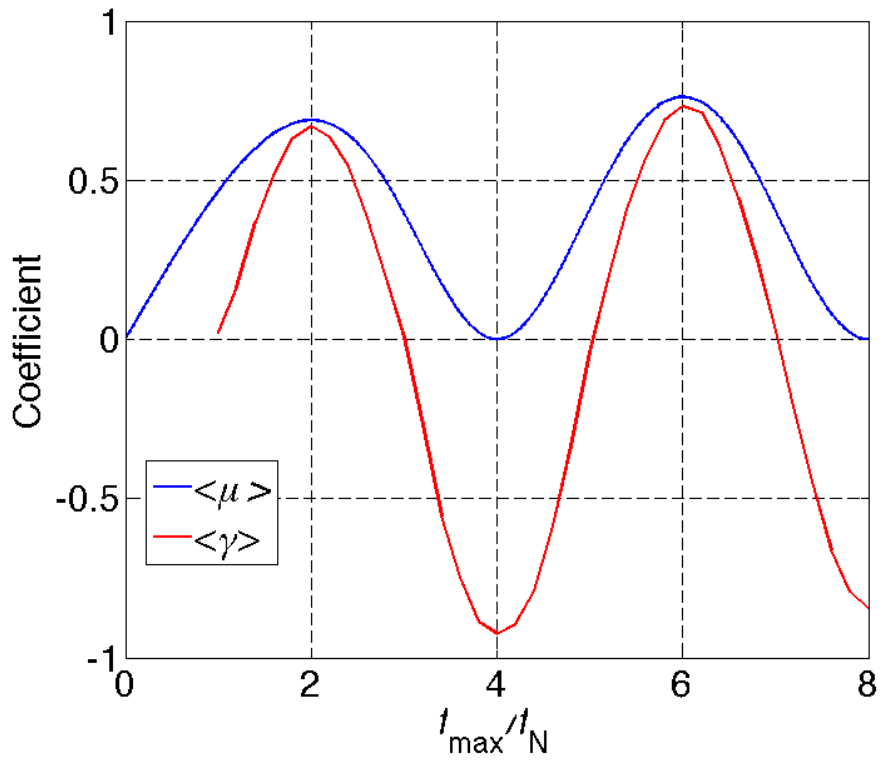

Fig. 3. Numerical evaluation of the parameters $\mu$ and $\gamma$ as a function of $f_{max}/f_N$ assuming the signal PSD given by Eq. (20). $\gamma$ is estimated by evaluating Eq. (27) for a number of independently generated data streams.

We now return to Eq. (16) and treat the more general case in which the electrical filter cutoff frequency $f_F$ can also be smaller than the maximum signal frequency $f_{max}$. However, we still assume $f_F \geq f_{max}/2$. The case $f_F < f_{max}/2$ results in further corrective terms and is cumbersome to derive while being of very limited practical relevance: In a typical optically amplified link, an optical filter is interposed between the optical amplifier and the photodetector. As further discussed in the next section, in the formalism used here this results in $f_{max}$ being smaller than the single sided optical filter bandwidth $f_{OF}$ (half the optical filter passband assuming the optical carrier to be centered relative to the filter's passband). $f_F < f_{max}/2$ would then correspond to the electrical filter cutoff frequency to be less than half the optical filter's cutoff frequency $f_{OF}$, which would be a poor system design choice due to the increased ASE-ASE beat noise arising from the unnecessary wide optical passband. With the assumption $f_F \geq f_{max}/2$ we convert Eq. (16) into

$$\sigma_{0/1} = \sqrt{N_n \cdot \left[\int_{f_n=0}^{f_n=f_F} \left|\frac{4\sqrt{P_{SOA,0/1}} \pm \int_{f_{s1}=f_n}^{f_{s1}=f_{max}} b^*\, df_{s1} \pm \int_{f_{s2}=2f_F-f_n}^{f_{s2}=f_{max}} b\, df_{s2}}{4}\right|^2 df_n + \int_{f_n=0}^{f_n=f_{max}} \left|\frac{\int_{f_{s3}=f_n}^{f_{s3}=f_{max}} b\, df_{s3}}{4}\right|^2 df_n\right]} \quad (29)$$

Additional difficulties arise here from the fact that the upper integration bound of one of the integrals in Eq. (16), $2f_F - f_n$, cannot be simply set to $f_{max}$, resulting in an additional term in Eq. (29) as compared to Eq. (17). Note that the subscripts in $df_{s1}$, $df_{s2}$ and $df_{s3}$ are introduced to help mapping the terms to the next equation. Eq. (29) is developed into a similar expression as Eq. (18) with a few additional terms

$$\sigma_{0/1}{}^2/N_n = P_{SOA,0/1}f_F \pm \frac{1}{2}\sqrt{P_{SOA,0/1}}\left(\int_{f_{s1}=0}^{f_{s1}=f_{max}} Re(b) min(f_{s1}, f_F)\, df_{s1} + \int_{f_{s2}=f_F}^{f_{s2}=f_{max}} Re(b)(f_{s2} - f_F)\, df_{s2}\right)$$
$$+\frac{1}{16}\left(\int_{f_{s1a}=0}^{f_{s1a}=f_{max}} \int_{f_{s1b}=0}^{f_{s1b}=f_{max}} min(f_{s1a}, f_{s1b}, f_F)\left(Re\left(b(f_{s1a})\right)Re\left(b(f_{s1b})\right) + Im\left(b(f_{s1a})\right)Im\left(b(f_{s1b})\right)\right) df_{s1a}df_{s1b}\right) \quad (30)$$

$$+\frac{1}{16}\left(\int_{f_{s2a}=f_F}^{f_{s2a}=f_{max}}\int_{f_{s2b}=f_F}^{f_{s2b}=f_{max}} min(f_{s2a}-f_F,f_{s2b}-f_F)\left(Re\big(b(f_{s2a})\big)Re\big(b(f_{s2b})\big) + Im\big(b(f_{s2a})\big)Im\big(b(f_{s2b})\big)\right)df_{s2a}df_{s2b}\right)$$

$$+\frac{2}{16}\left(\int_{f_{s1}=0}^{f_{s1}=f_F}\int_{f_{s2}=f_F}^{f_{s2}=f_{max}} max(f_{s1}+f_{s2}-2f_F,0)\left(Re\big(b(f_{s1})\big)Re\big(b(f_{s2})\big)-Im\big(b(f_{s1})\big)Im\big(b(f_{s2})\big)\right)df_{s1}df_{s2}\right)$$

$$+\frac{2}{16}\left(\int_{f_{s1}=f_F}^{f_{s1}=f_{max}}\int_{f_{s2}=f_F}^{f_{s2}=f_{max}}(f_{s2}-f_F)\left(Re\big(b(f_{s1})\big)Re\big(b(f_{s2})\big)-Im\big(b(f_{s1})\big)Im\big(b(f_{s2})\big)\right)df_{s1}df_{s2}\right)$$

$$+\frac{1}{16}\left(\int_{f_{s3a}=0}^{f_{s3a}=f_{max}}\int_{f_{s3b}=0}^{f_{s3b}=f_{max}} min(f_{s3a},f_{s3b})\left(Re\big(b(f_{s3a})\big)Re\big(b(f_{s3b})\big)+Im\big(b(f_{s3a})\big)Im\big(b(f_{s3b})\big)\right)df_{s3a}df_{s3b}\right)$$

Since

$$\int_{f_{s1}=0}^{f_{s1}=f_{max}} Re(b)min(f_{s1},f_F)\,df_{s1}+\int_{f_{s2}=f_F}^{f_{s2}=f_{max}} Re(b)(f_{s2}-f_F)\,df_{s2}=\int_{f_s=0}^{f_s=f_{max}} Re(b)f_s\,df_s \tag{31}$$

and

$$\begin{aligned}&\int_{f_{s1a}=0}^{f_{s1a}=f_{max}}\int_{f_{s1b}=0}^{f_{s1b}=f_{max}} min(f_{s1a},f_{s1b},f_F)\left(Re\big(b(f_{s1a})\big)Re\big(b(f_{s1b})\big)+Im\big(b(f_{s1a})\big)Im\big(b(f_{s1b})\big)\right)df_{s1a}df_{s1b}\\&+\int_{f_{s2a}=f_F}^{f_{s2a}=f_{max}}\int_{f_{s2b}=f_F}^{f_{s2b}=f_{max}} min(f_{s2a}-f_F,f_{s2b}-f_F)\left(Re\big(b(f_{s2a})\big)Re\big(b(f_{s2b})\big) + Im\big(b(f_{s2a})\big)Im\big(b(f_{s2b})\big)\right)df_{s2a}df_{s2b}\\&=\int_{f_{s1a}=0}^{f_{s1a}=f_{max}}\int_{f_{s1b}=0}^{f_{s1b}=f_{max}} min(f_{s1a},f_{s1b})\left(Re\big(b(f_{s1a})\big)Re\big(b(f_{s1b})\big)+Im\big(b(f_{s1a})\big)Im\big(b(f_{s1b})\big)\right)df_{s1a}df_{s1b}\end{aligned} \tag{32}$$

the only modification relative to Eqs. (18) and (28) arises from the two terms

$$\begin{aligned}&\frac{2}{16}\left(\int_{f_{s1}=0}^{f_{s1}=f_F}\int_{f_{s2}=f_F}^{f_{s2}=f_{max}} max(f_{s1}+f_{s2}-2f_F,0)\left(Re\big(b(f_{s1})\big)Re\big(b(f_{s2})\big)-Im\big(b(f_{s1})\big)Im\big(b(f_{s2})\big)\right)df_{s1}df_{s2}\right)\\&+\frac{2}{16}\left(\int_{f_{s1}=f_F}^{f_{s1}=f_{max}}\int_{f_{s2}=f_F}^{f_{s2}=f_{max}}(f_{s2}-f_F)\left(Re\big(b(f_{s1})\big)Re\big(b(f_{s2})\big)-Im\big(b(f_{s1})\big)Im\big(b(f_{s2})\big)\right)df_{s1}df_{s2}\right)\\&=\frac{1}{16}\left(\int_{f_{s1}=0}^{f_{s1}=f_F}\int_{f_{s2}=f_F}^{f_{s2}=f_{max}} max(f_{s1}+f_{s2}-2f_F,0)\left(Re\big(b(f_{s1})\big)Re\big(b(f_{s2})\big)-Im\big(b(f_{s1})\big)Im\big(b(f_{s2})\big)\right)df_{s1}df_{s2}\right)\\&+\frac{1}{16}\left(\int_{f_{s1}=f_F}^{f_{s1}=f_{max}}\int_{f_{s2}=0}^{f_{s2}=f_F} max(f_{s1}+f_{s2}-2f_F,0)\left(Re\big(b(f_{s1})\big)Re\big(b(f_{s2})\big)-Im\big(b(f_{s1})\big)Im\big(b(f_{s2})\big)\right)df_{s1}df_{s2}\right)\\&+\frac{1}{16}\left(\int_{f_{s1}=f_F}^{f_{s1}=f_{max}}\int_{f_{s2}=f_F}^{f_{s2}=f_{max}}(f_{s1}+f_{s2}-2f_F)\left(Re\big(b(f_{s1})\big)Re\big(b(f_{s2})\big)-Im\big(b(f_{s1})\big)Im\big(b(f_{s2})\big)\right)df_{s1}df_{s2}\right)\\&=\frac{1}{16}\left(\int_{f_{s1}=0}^{f_{s1}=f_{max}}\int_{f_{s2}=2f_F-f_{s1}}^{f_{s2}=f_{max}}(f_{s1}+f_{s2}-2f_F)\left(Re\big(b(f_{s1})\big)Re\big(b(f_{s2})\big)-Im\big(b(f_{s1})\big)Im\big(b(f_{s2})\big)\right)df_{s1}df_{s2}\right)\end{aligned} \tag{33}$$

Applying a change of variable $f=f_{s1}+f_{s2}$ we transform this expression into

$$\int_{f_{s1}=-f_{max}}^{f_{s1}=f_{max}}\int_{f=2f_F}^{f=2f_{max}}(f-2f_F)\left(\frac{Re(b(f_{s1}))}{4}\frac{Re(b(f-f_{s1}))}{4}-\frac{Im(b(f_{s1}))}{4}\frac{Im(b(f-f_{s1}))}{4}\right)df_{s1}df$$
$$=\int_{f=2f_F}^{f=2f_{max}}(f-2f_F)Re\left(F\left(\left(\sqrt{P_{SOA}}-mean(\sqrt{P_{SOA}})\right)^2\right)\right)df \tag{34}$$

In other words, Eq. (28) remains valid, provided we replace $\gamma$ by

$$\bar{\gamma}f_N=\frac{8\int_{f=0}^{f=\infty}\int_{f_{s1}=-\infty}^{f_{s1}=\infty}min(f,2f_F)\left(\frac{Re(b(f_{s1}))}{4}\frac{Re(b(f-f_{s1}))}{4}-\frac{Im(b(f_{s1}))}{4}\frac{Im(b(f-f_{s1}))}{4}\right)df_{s1}df}{\left(\sqrt{P_{SOA,1}}-\sqrt{P_{SOA,0}}\right)^2}$$
$$=\frac{8\int_{f=0}^{f=2f_{max}}min(f,2f_F)Re\left(F\left(\left(\sqrt{P_{SOA}}-mean(\sqrt{P_{SOA}})\right)^2\right)\right)df}{\left(\sqrt{P_{SOA,1}}-\sqrt{P_{SOA,0}}\right)^2} \tag{35}$$

$\mu$ on the other hand remains unchanged. As seen in Fig. 4, the effect of the electrical filter bandwidth on the coefficient $\bar{\gamma}$ remains small even in the extreme case given by $f_F=f_{max}/2$.

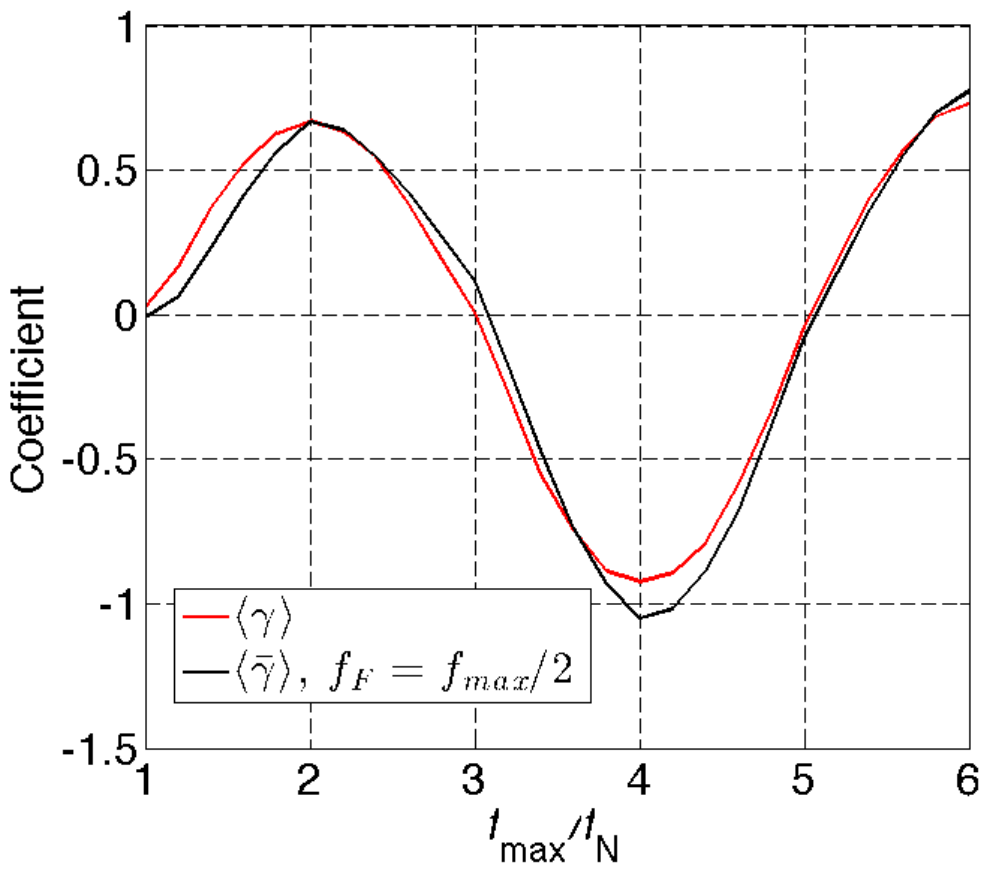

Fig. 4. Comparison between the coefficient $\gamma$ calculated for $f_F>f_{max}$ and the coefficient $\bar{\gamma}$ calculated for $f_F=f_{max}/2$. The difference can be seen to be slight.

## 4. Generalization to the case of optical and electrical filtering

We now assume that an optical filter with a passband of $2f_{OF}$ has been interposed between the SOA and the Rx (corresponding to filtering of the data with a single sided cutoff frequency $f_{OF}$). The case $f_{OF}\geq f_F$ is relatively straightforward to analyze, given the above, since the integration domain of the second term of Eq. (14) simply needs to be restricted resulting in

$$\sigma_{0/1}=\sqrt{N_n\cdot\left[\int_{f_n=0}^{f_n=f_F}\left|\frac{4\sqrt{P_{SOA,0/1}}\pm\int_{f_s=f_F-f_n}^{f_s=f_{max}}2Re(b)\,df_s\mp\int_{f_s=f_F-f_n}^{f_s=f_F+f_n}b\,df_s}{4}\right|^2df_n+\int_{f_n=f_F}^{f_n=min(f_{max}+f_F,f_{OF})}\left|\frac{\int_{f_s=f_n-f_F}^{f_s=f_n+f_F}b\,df_s}{4}\right|^2df_n\right]} \tag{36}$$

Since the derivation of $\sigma_{0/1}$ is based on the signal Fourier components that reach the Rx, $P_{SOA,0/1}$ refers to the signal levels after optical filtering referred back to the input of the SOA. In other words, signal distortion due to ISI associated with the optical filtering is applied to the $P_{SOA,0/1}$ levels. They are obtained by taking the signal levels after the optical filter and dividing them by the gain of the SOA. Moreover, $f_{max}$ (as also used to evaluate the expectation values of $\mu$ and $\gamma$) refers to the maximum signal frequency after optical filtering and is equal to $f_{OF}$ is it was initially higher than the latter.

If $f_{OF}\geq f_F+f_{max}$, the optical filter has no effect on the ASE-signal beat noise levels and we can revert to the formulas from the previous section. Thus, assuming $f_F\leq f_{OF}\leq f_F+f_{max}$ we obtain

$$\sigma_{0/1} = \sqrt{N_n \cdot \left[ \int_{f_n=0}^{f_n=f_F} \left| \frac{4\sqrt{P_{SOA,0/1}} \pm \int_{f_s=f_F-f_n}^{f_s=f_{max}} 2Re(b)\, df_s \mp \int_{f_s=f_F-f_n}^{f_s=f_F+f_n} b df_s}{4} \right|^2 df_n + \int_{f_n=0}^{f_n=f_{OF}-f_F} \left| \frac{\int_{f_s=f_n}^{f_s=f_n+2f_F} b\, df_s}{4} \right|^2 df_n \right]} \quad (37)$$

Further assuming, as previously, $f_F \geq f_{max}/2$ we obtain

$$\sigma_{0/1} = \sqrt{N_n \cdot \left[ \int_{f_n=0}^{f_n=f_F} \left| \frac{4\sqrt{P_{SOA,0/1}} \pm \int_{f_s=f_F-f_n}^{f_s=f_{max}} 2Re(b)\, df_S \mp \int_{f_s=f_F-f_n}^{f_s=f_F+f_n} b df_s}{4} \right|^2 df_n + \int_{f_n=0}^{f_n=f_{OF}-f_F} \left| \frac{\int_{f_s=f_n}^{f_s=f_{max}} b\, df_s}{4} \right|^2 df_n \right]} \quad (38)$$

Only the second term inside the square root differs from the previous derivations, so that we can directly focus on the latter. If $f_{OF} = f_F$ Eq. (28) can be simply updated into

$$\begin{aligned} \sigma_{0/1}{}^2/N_n &= P_{SOA,0/1} f_F \pm \frac{1}{2}\sqrt{P_{SOA,0/1}}\left(\sqrt{P_{SOA,1}} - \sqrt{P_{SOA,0}}\right)\mu f_N + \frac{1}{16}\left(\sqrt{P_{SOA,1}} - \sqrt{P_{SOA,0}}\right)^2 (2\mu - \gamma) f_N \\ &= P_{SOA,0/1} f_F \pm \frac{1}{4}\left(P_{SOA,1} - P_{SOA,0}\right)\mu f_N - \frac{1}{8}\left(\sqrt{P_{SOA,1}} - \sqrt{P_{SOA,0}}\right)^2 \left(\mu + \frac{\gamma}{2}\right) f_N \end{aligned} \quad (39)$$

The multiplicative factor $1/8$ in front of the third term is converted to $1/16$ since one of the two contributing terms (that are equal to each other when $f_{max} < f_F$) is filtered out. $\bar{\gamma} = \gamma$ since after optical filtering $f_{max} \leq f_{OF} = f_F$ (if $f_{max}$ was originally larger than $f_{OF}$, after optical filtering the two are equal). For intermediate cases the contribution from the second term of Eq. (38) needs to be restricted as

$$\sigma_{0/1}{}^2/N_n = \ldots + \frac{1}{16}\left( \int_{f_{s1}=0}^{f_{s1}=f_{max}} \int_{f_{s2}=0}^{f_{s2}=f_{max}} min(f_{s1}, f_{s2}, f_{OF} - f_F) \left( Re\big(b(f_{s1})\big) Re\big(b(f_{s2})\big) + Im\big(b(f_{s1})\big) Im\big(b(f_{s2})\big) \right) df_{s1} df_{s2} \right) \quad (40)$$

So that the corrective term to be applied to Eqs. (28) is

$$-\frac{1}{16}\left( \int_{f_{s1}=f_{OF}-f_F}^{f_{s1}=f_{max}} \int_{f_{s2}=f_{OF}-f_F}^{f_{s2}=f_{max}} min(f_{s1} - f_{OF} + f_F, f_{s2} - f_{OF} + f_F) \left( Re\big(b(f_{s1})\big) Re\big(b(f_{s2})\big) + Im\big(b(f_{s1})\big) Im\big(b(f_{s2})\big) \right) df_{s1} df_{s2} \right) \quad (41)$$

On the other hand, if $f_{OF} < f_F$ Eq. (36) reduces to

$$\sigma_{0/1} = \sqrt{N_n \cdot \left[ \int_{f_n=0}^{f_n=f_{OF}} \left| \frac{4\sqrt{P_{SOA,0/1}} \pm \int_{f_s=f_F-f_n}^{f_s=f_{max}} 2Re(b)\, df_S \mp \int_{f_s=f_F-f_n}^{f_s=f_F+f_n} b df_s}{4} \right|^2 df_n \right]} \quad (42)$$

Since in this case $f_{max} \leq f_{OF} \leq f_F$ also holds, this further simplifies into

$$\sigma_{0/1} = \sqrt{N_n \cdot \left[ \int_{f_n=0}^{f_n=f_{OF}} \left| \frac{4\sqrt{P_{SOA,0/1}} \pm \int_{f_s=f_F-f_n}^{f_s=f_{max}} b^*\, df_s}{4} \right|^2 df_n \right]} \quad (43)$$

Which results exactly in Eq. (39) with the difference that $f_F$ has to be replaced by $f_{OF}$, i.e.,

$$\sigma_{0/1}{}^2/N_n = P_{SOA,0/1} f_{OF} \pm \frac{1}{4}\left(P_{SOA,1} - P_{SOA,0}\right)\mu f_N - \frac{1}{8}\left(\sqrt{P_{SOA,1}} - \sqrt{P_{SOA,0}}\right)^2 \left(\mu + \frac{\gamma}{2}\right) f_N \quad (44)$$

As already stated, the values $P_{SOA,0/1}$ are modified according to ISI occurring due to the optical filtering (electrical filtering is not taken into account in $P_{SOA,0/1}$, as these correspond to the signal shape entering the Rx and divided by the SOA gain).

When an optical filter is interposed between the SOA and the Rx, one subtlety arises from the fact that the signal levels $P_{SOA,0/1}$ depend on the details of the bit sequence, since the ISI induced by the optical filter is input referred to the SOA. Strictly, Eq. (44) should thus be recast in a form where the signal levels are attributable to their instantaneous values, i.e.,

$$\sigma_0{}^2/N_n = P_{SOA,0}f_{OF} + \frac{a}{2}\left(\frac{a}{2} - \sqrt{P_{SOA,0}}\right)\mu f_N - \frac{1}{2}\left(\frac{a}{2} - \sqrt{P_{SOA,0}}\right)^2\left(\mu + \frac{\gamma}{2}\right)f_N$$

$$\sigma_1{}^2/N_n = P_{SOA,1}f_{OF} - \frac{a}{2}\left(\sqrt{P_{SOA,1}} - \frac{a}{2}\right)\mu f_N - \frac{1}{2}\left(\sqrt{P_{SOA,1}} - \frac{a}{2}\right)^2\left(\mu + \frac{\gamma}{2}\right)f_N \qquad (45)$$

i.e.,

$$\sigma_{0/1}{}^2/N_n = P_{SOA}f_{OF} - \frac{a}{2}\left(\sqrt{P_{SOA}} - \frac{a}{2}\right)\mu f_N - \frac{1}{2}\left(\sqrt{P_{SOA}} - \frac{a}{2}\right)^2\left(\mu + \frac{\gamma}{2}\right)f_N$$

These expressions are justified by the fact that the coefficients $\sqrt{P_{SOA,0/1}}$ arise from an integration of $b(\omega)$ and thus correspond to the instantaneous value of the signal strength. Once a substantial level of ISI is assumed, further refinements would need to be taken into account for a fully rigorous derivation, such as the fact that $\left(\frac{d\sqrt{P_{SOA}}}{dt}\right)_{t=0} = 0$ does not hold anymore (the signal can have a finite slope at the sampling time in a highly distorted case). Moreover, the dependence of the coefficients $\mu$ and $\gamma$ on the data pattern might correlate with the instantaneous ISI penalty, which would have to be taken into account for the most accurate estimation of the signal Q-factor. Nonetheless, Eq. (45) has proven heuristically to adequately model situations with optical filter induced ISI as shown by the numerical validation reported in section 5, in which the data dependence of $\mu$ and $\gamma$ is also numerically investigated.

## 5. Numerical verification of models

In order to numerically verify the equations derived in the previous sections we run a series of numerical experiments corresponding to the different scenarios. The models are also compared to experimental data from ref. [4].

In the numerical experiments, a signal is first generated by sending a random bit sequence with ideal square shaped symbols through an ideal low pass filter (i.e., by truncating its Fourier transform) with a cutoff frequency $f_{max}$ expressed as a fraction of $f_N$ in the summary table shown below. Both the ASE noise, modeled as white noise, and the signal are filtered by an optical notch filter with a passband $2f_{OF}$. The ASE noise is then multiplied with the signal amplitude in order to model signal-ASE beat noise. Finally, the modulated ASE-signal beat noise is sent through an electrical low pass filter with a cutoff frequency $f_F$. After sampling of the noise at the signal sampling times the std of the 0- and 1-level noise ($\sigma_0$ and $\sigma_1$) is extracted.

As a reference, noise levels are also calculated by feeding the noise through a filter with a single sided cutoff $f_{OF}$ or $f_F$ (whichever of the two is the smaller number) followed by multiplication with the 0- and 1-bit signal amplitude *after* filtering. These are the noise levels $\overline{\sigma_0}$ and $\overline{\sigma_1}$ corresponding to the instantaneous ASE-beat noise levels in the limit of very slowly varying signals.

Two numbers are then extracted from the level dependent noise std: One corresponds to the transfer of noise between the variance of the 1-level and the variance of the 0-level noise. The other corresponds to the net increase or decrease of noise seen in the sum of the two variances. These two numbers are defined as

$$\beta_{diff} = -2\frac{(\sigma_1^2 - \sigma_0^2) - (\bar{\sigma}_1^2 - \bar{\sigma}_0^2)}{(\bar{\sigma}_1^2 - \bar{\sigma}_0^2)}\frac{min(f_F, f_{OF})}{f_N} \qquad (46)$$

$$\beta_{comm} = -4\frac{(\sigma_1^2 + \sigma_0^2) - (\bar{\sigma}_1^2 + \bar{\sigma}_0^2)}{(\overline{\sigma_1} - \overline{\sigma_0})^2}\frac{min(f_F, f_{OF})}{f_N} \qquad (47)$$

The first term is expected to be equal to $\mu$ and the second term is expected to be equal to $\gamma$, $\bar{\gamma}$ or $\mu + \gamma/2$, respectively when Eqs. (28), (35) or (44) apply. The results are summarized in the following table (numeric data corresponds to the average recorded over 10000 randomly generated data streams of 300 bits each):

Table 1: Comparison of noise mixing coefficients expected from the numerical evaluation of the analytical expressions with the coefficients extracted from a numerically modelled data stream.

| $f_{max}$ | $f_F$ | $f_{OF}$ | Applicable Formula | $\beta_{diff}$ Numeric | $\mu$ Analytical | $\beta_{comm}$ Numeric | $\beta_{comm}$ Analytical* |
|---|---|---|---|---|---|---|---|
| $2f_N$ | $f_N$ | $\infty$ | Eqs. (21), (35), (28) | 0.6984 | 0.6881 | 0.7001 | 0.67 |
| $2f_N$ | $2f_N$ | $\infty$ | Eqs. (21), (27), (28) | 0.6922 | 0.6881 | 0.6989 | 0.67 |
| $2f_N$ | $4f_N$ | $\infty$ | Eqs. (21), (27), (28) | 0.6922 | 0.6881 | 0.6995 | 0.67 |
| $4f_N$ | $2f_N$ | $\infty$ | Eqs. (21), (35), (28) | 0.0028 | 0 | -1.0030 | -1.05 |
| $4f_N$ | $4f_N$ | $\infty$ | Eqs. (21), (27), (28) | 0.0034 | 0 | -0.8939 | -0.92 |
| $4f_N$ | $6f_N$ | $\infty$ | Eqs. (21), (27), (28) | 0.0013 | 0 | -0.9030 | -0.92 |
| $2f_N$ | $2f_N$ | $2f_N$ | Eqs. (21), (27), (44) | 0.6915 | 0.6881 | 1.0589 | 1.03 |
| $2f_N$ | $4f_N$ | $2f_N$ | $f_F \geq f_{max} + f_N$ | 0 | 0 | 0 | 0 |
| $4f_N$ | $4f_N$ | $4f_N$ | Eqs. (21), (27), (44) | 0.0039 | 0 | -0.4416 | -0.46 |

*The analytical formulas Eqs. (27) and (35) are evaluated based on the Fourier transform of random data streams, with results shown in Figs. 3 and 4, while the analytical formula Eq. (21) is directly evaluated as the numerical evaluation of a simple integral.

The numerical results shown in Table 1 validate the analytical formulas derived in the previous sections. One shortcoming of the derived formulas remains however in the fact that the coefficients relate to the changes applied to the averaged 0- and 1-level noise. However, as already discussed above, the actual noise levels do not only depend on whether the bit is a ‘0’ or a ‘1’, but also on the sequence of bits preceding and succeeding the detected bit. We numerically investigate the pattern dependency in further details in the the situation $f_{OF} = f_F \leq f_{max}$ for different values of $f_{OF}/f_N$ and investigate the correlation of the noise levels with the ISI penalized signal levels, as well as with the preceding and succeeding bit sequences. Each bit is classified into one of three categories: Category 1 corresponds to bits for which both the immediately preceding and the immediately succeeding bit are different, i.e., the center bit of a 010 or a 101 bit sequence. Category 2 corresponds to bits for which only one out of the immediately preceding or the immediately succeeding bit is different, i.e., the center bit of a 001, 110, 011 or 100 sequence. Category 3 corresponds to bits for which both adjacent bits are identical to them, i.e., the center bit if a 000 or 111 sequence.

Figure 5 shows the 0- and 1-level noise as a function of the signal levels for different $f_{OF}/f_N$. The classification of each bit is indicated by color coding. The data from the numerical model is overlaid with the predictions from Eq. (45), which applies in this case, as well as a prediction based on the simple value of the ASE-signal beat noise derived in the limit of a slowly varying signal ($\overline{\sigma_{0/1}}$).

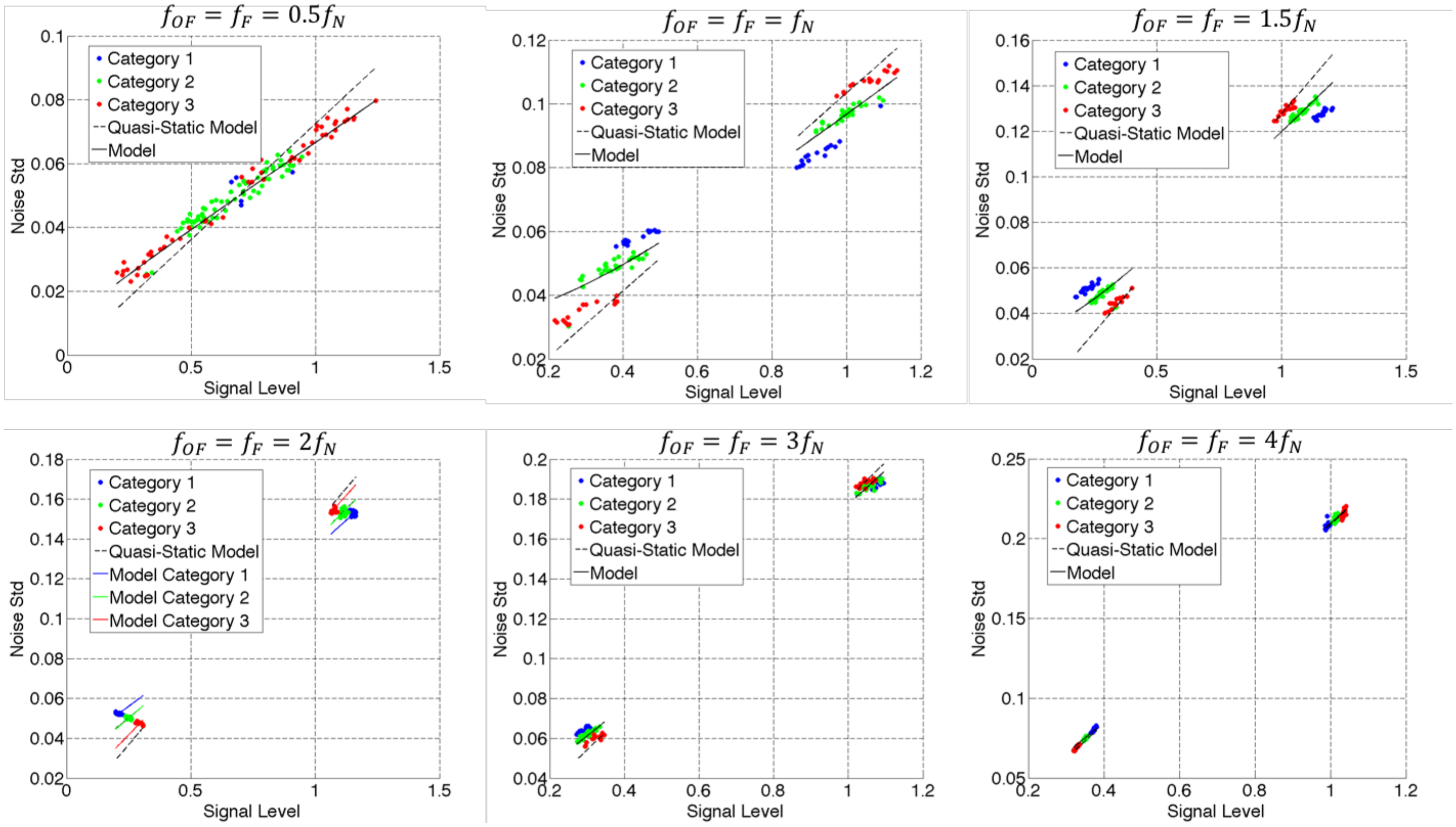

Fig. 5. Std of optically and electrically filtered ASE-signal beat noise extracted from a numerical model for different choices of $f_{OF} = f_F$. Each dot represents a bit of a PRBS-7 signal. Dots are color coded according to the classification of the bits into one of the three bit categories. The noise std is plotted as a function of the corresponding sampled signal level. The dashed line corresponds to an ASE-signal beat noise model derived in the limit of a slowly varying signal, i.e., corresponding to $\overline{\sigma_0}$ and $\overline{\sigma_1}$, that does not take noise mixing into account. The continuous black line corresponds to the model given by Eq. (45). For the case $f_{OF} = f_F = 2f_N$ the three categories of bits have been individually modelled and the coefficients $\mu$ and $\gamma$ recalculated for each of them (model shown by colored lines).

Following observations can be made in Fig. 5:

- For the case $f_{OF} = f_F = 0.5f_N$ the std of the ASE-signal beat noise is adequately predicted by Eq. (45) together with the average coefficients $\langle\mu\rangle$ and $\langle\gamma\rangle$. The noise std follows the dependence on the SOA input power level ($P_{SOA}$) predicted by Eq. (45). Here, ISI is very high (the eye diagram is fully closed) and drives both the variation of sampled signal levels as well as recorded ASE noise levels.
- For the cases $f_{OF} = f_F = 1f_N$ and $f_{OF} = f_F = 1.5f_N$ with less extreme ISI, the predicted dependence of the ASE noise on the signal level can still be globally seen (and can be very well seen within each category of bits), however an increasingly clear separation between the distributions corresponding to distinct categories of bits can be seen.
- In the case $f_{OF} = f_F = 2f_N$ the bit category can be seen to be the driving factor determining the noise std.
- Finally, in the case $f_{OF} = f_F = 4f_N$ the std of all the bits return to the distribution derived in the limit of slowly varying signals, related to the fact that in this case $\langle\mu\rangle = 0$.

The different signal level dependencies of the noise std as seen for different bit categories are due to the fact that the bit specific coefficients $\mu$ and $\gamma$ depend on the category. This was verified by plotting the coefficients $\mu$ and $\gamma$ extracted from a numerically modeled data stream by using Eqs. (46) and (47). The coefficients were extracted for all the bits of a pseudo random PRBS-7 bit sequence for the exemplary situation $f_{OF} = f_F = 2f_N$ and subsequently classified by bit category (Fig. 6). It can be seen that the bit categorization in terms of their nearest neighbors coincides with the clustering of the coefficients in three clearly distinguishable clusters. The coefficients $\mu$ and $\gamma$ were then averaged for each bit category with the results summarized below for $f_{OF} = f_F = 2f_N$:

Table 2: Average noise mixing coefficient for each category of bit in the case $f_{OF} = f_F = 2f_N \leq f_{max}$.

| | Category 1 | Category 2 | Category 3 |
|---|---|---|---|
| $\mu$ | 1.05 | 0.69 | 0.24 |
| $\gamma$ | 0.99 | 0.71 | 0.27 |

The noise std vs. signal level was then independently modeled according to Eq. (45) for each bit category and plotted in the corresponding graph of Fig. 5. It can be seen that the category specific models predict the noise std very well and adequately account for pattern dependent effects.

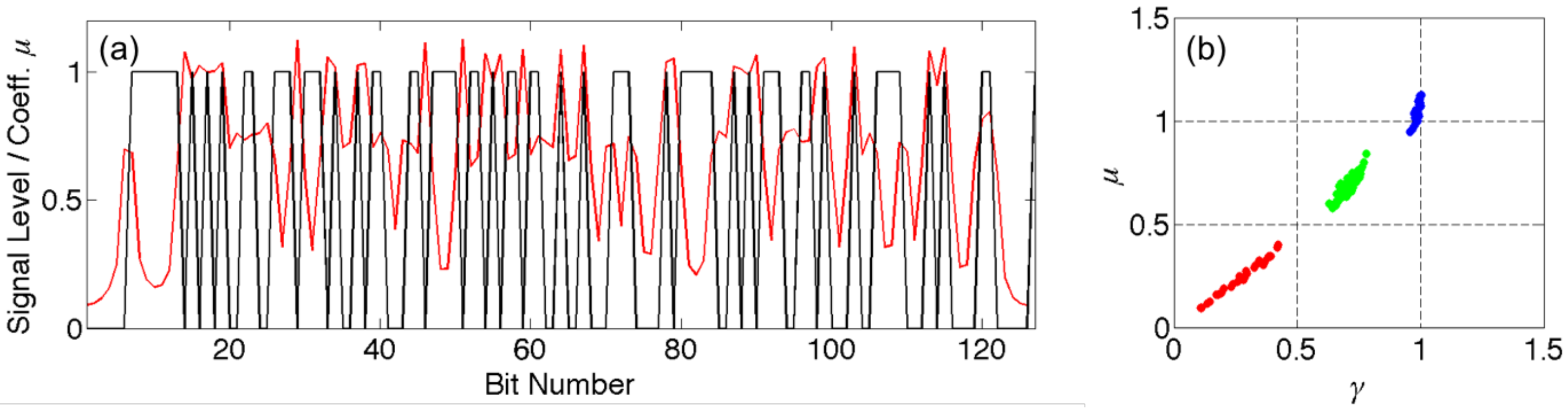

Fig. 6. (a) Overlay of the bit sequence (black) with the extracted bit-specific coefficient $\mu$ (red). (b) Bit specific coefficients $\mu$ and $\gamma$ extracted from the PRBS-7 sequence with $f_{OF} = f_F = 2f_N$. Three clusters are clearly visible that coincide with the bit categories. Category 1 bits are shown in blue, category 2 bits are shown in green and category 3 bits are shown in red.

As expected, category 1 bits have the highest coefficients $\mu$ and $\gamma$ as they correspond to fast switching bit sequences with the highest high frequency signal frequency content (as explained in section 2, these effects increase with signal frequency). Conversely, category 3 bits have the smallest coefficients as they correspond to the least amount of nearby signal transitions. Interestingly, the coefficients for category 1 bits are close to 1, as expected when approximation a 010101… bit sequence by a sine wave oscillating at the Nyquist frequency (see section 2). The simple derivation of section 2 is thus shown to retain a high degree of relevance, as it corresponds to the bits limiting the BER both in terms of vertical eye opening (in a typical low pass filtered data stream system, not necessarily true when effects such as amplifier or modulator peaking [5] or SOA saturation [4] play a role) and in terms of ASE-signal beat noise.

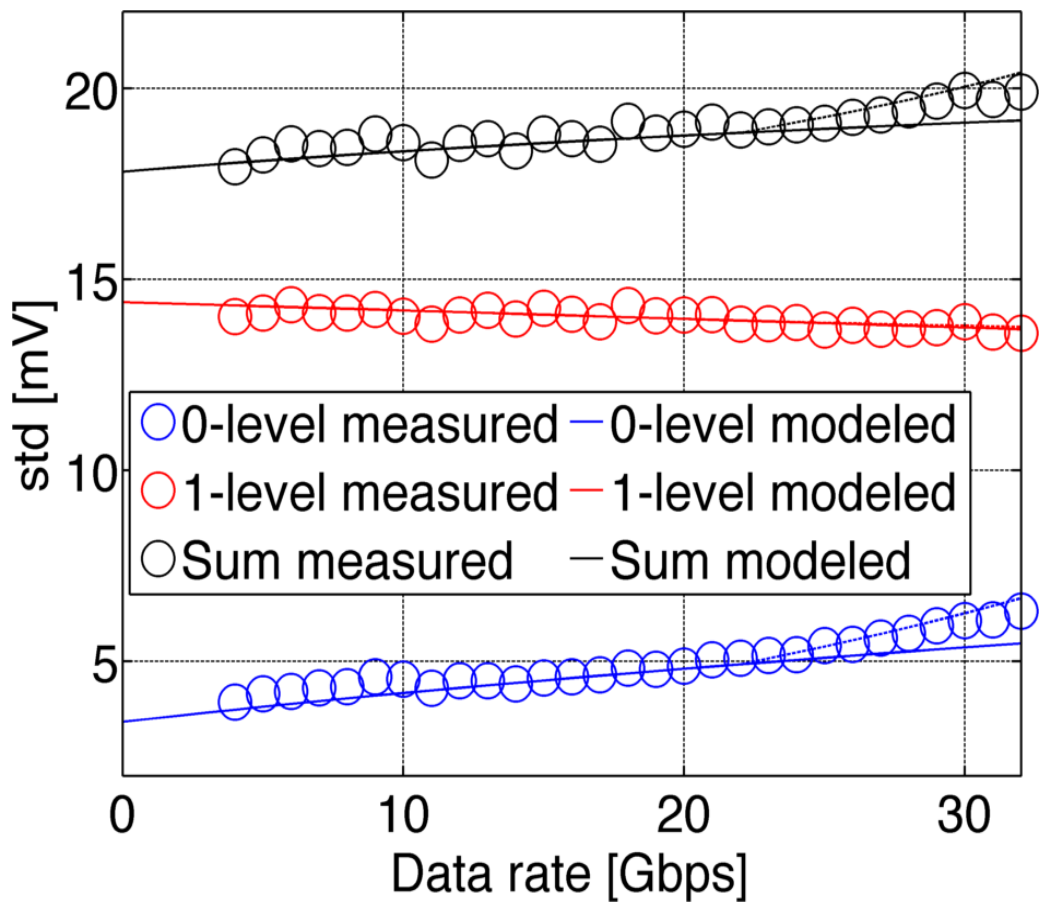

Fig. 7. Experimental verification of the model described by Eq. (44). The dots show the measured values of the 0- and 1-level noise std, as well as their sum (the denominator of the signal Q-factor). The continuous lines correspond to the model given by Eq. (44), but without taking ISI induced by the optical filter into account. The doted lines take into account a progressive reduction of the signal extinction from 13 dB to 10 dB occurring for data rates between 20 and 32 Gbps due to signal clipping by the optical filter as well as the finite cutoff frequency of the optical modulator. The reduced extinction occurring prior to opto-electric transduction further penalizes the 0-level ASE-signal beat noise.

In order to experimentally verify the model given by Eq. (44), we modulated a -14 dBm optical carrier with a commercial Mach-Zehnder Modulator (MZM) with a 33.5 GHz cutoff frequency and amplified it with an SOA located downstream of the modulator and followed by a 40 GHz passband optical filter ($f_{OF} \simeq 20$ GHz). After being converted back to the electrical domain with a commercial high-speed photo-receiver (U2T/Finisar XPRV2021A) with a 40 GHz bandwidth, the signal was recorded for data rates between 4 and 32 Gbps with a real time oscilloscope with an analog bandwidth ($f_F$) of 21 GHz (corresponding to an approximately square electrical filter due to the steep higher order roll off). The std of the 1- and 0-levels was extracted and is shown in Fig. 7 (dots). The continuous lines show the predictions based on Eq. (44) with coefficient $\mu$ and $\gamma$ extracted from a numerical simulation of the data stream taking the experimentally recorded waveform into account. The dashed lines further take into account the reduced signal extinction occurring at high data rates due to ISI induced by the optical filter. Increased effective noise levels, as given by $\sigma_0 + \sigma_1$, are in particular clearly apparent in the experimental data. This trend would not have been expected with a simpler noise model, as the noise equivalent bandwidth of the Rx stayed unchanged.

## 6. Different link configurations and relative intensity noise

We start this section by a discussion on how to treat the case when a booster optical amplifier is interposed between a laser operated in CW mode and a modulator, rather than a linear amplifier interposed between the modulator and the Rx. $P_{SOA}$ is then simply the constant optical power entering the SOA. We introduce the coefficients $T_{MOD,0}$ and $T_{MOD,1}$ defined as the power at the output of the modulator divided by the power at the input of the modulator, respectively for the 0- and 1-states. We further redefine the coefficients $a$ and $b(f_s)$ as

$$\sqrt{P_{SOA}}T_{MOD} = \frac{a}{2} + \int_{-\infty}^{\infty} \frac{b(f_s)}{4} e^{i2\pi f_s t} df_s \tag{48}$$

The formulas derived in the previous sections then remain valid, provided $\sqrt{P_{SOA,0/1}}$ is replaced by $\sqrt{P_{SOA}}T_{MOD,0/1}$.The analysis is not substantially changed by this modified link configuarion, as it depends on the signal spectrum at the entrance of the Rx.

It is further instructive to analyze the situation when laser RIN rather than ASE is the source of noise. A fundamental difference here is that while ASE noise is broadband and ASE-signal beat noise can be approximated as white noise for a constant signal in the absence of filtering, RIN has a strong spectral dependency and rolls off beyond the laser relaxation oscillation frequency. We change the notation $f_{max}$ into $f_{s,max}$ to explicitly identify it as the maximum signal frequency and introduce the notation $f_{n,max}$ defined as the maximum noise frequency (the RIN decays to the shot noise level at best, so that technically there is no maximum noise frequency, but we assume here that contributions of RIN to the noise budget are dominated by the low speed "classical" RIN). If the electrical filter bandwidth is over specified such that $f_F \geq f_{s,max} + f_{n,max}$, the electrical filtering has no effect on the level dependent RIN (even after up- and down conversion by multiplication with the signal, the noise spectrum remains fully within the passband of the electrical filter). Even if $f_F < f_{s,max} + f_{n,max}$, only signal frequency components verifying $f_s > f_F - f_{n,max}$ contribute to "mix" the noise levels (i.e., transfer some of the noise variance between the 0- and 1- signal levels as previously shown). In a situation where the RIN rolls off at a frequency $f_{n,max}$ that is substantially smaller than the electrical filter bandwidth and for a functional communication system in which most of the signal PSD falls within the filter bandwidth, only a small fraction of the signal PSD contributes, so that this effect remains very small and can be safely neglected. This would for example be the case in an externally modulated link in which the laser relaxation oscillation frequency is substantially below the Nyquist frequency of the data stream (such as in [4]). Conversely, in a directly modulated link in which signal frequency components are primarily below the laser relaxation frequency, "noise mixing" might also play a role in relation to RIN. In this case a similar analysis as previously applies, provided $\sqrt{N_n \cdot P_{SOA}}$ is also replaced by $\sqrt{N_n} \cdot T_{MOD,0/1}$ in the analysis.

## 7. Summary and conclusions

In summary, when sending an optical data stream through an optical amplifier prior to opto-electric transduction, level dependent ASE-signal beat noise as transformed by filtering in the electrical domain can be modeled as

$$\sigma_{0/1} = \sqrt{2G^2Fhf_0 \cdot \left[P_{SOA,0/1}f_F \pm \frac{1}{4}\left(P_{SOA,1} - P_{SOA,0}\right)\mu f_N - \frac{1}{8}\left(\sqrt{P_{SOA,1}} - \sqrt{P_{SOA,0}}\right)^2 \gamma f_N\right]}$$

where $G$ and $F$ are the gain and noise factor of the optical amplifier, $h$ is Planck's constant, $f_0$ is the carrier frequency, $P_{SOA,0/1}$ are the 0- and 1-bit power levels entering the amplifier, $f_F$ is the noise equivalent bandwidth of the Rx and $f_N$ is the Nyquist frequency. The coefficients $\mu$ and $\gamma$ depend on the signal shape

$$\sqrt{P_{SOA}} = \frac{a}{2} + \int_{-\infty}^{\infty} \frac{b(f_s)}{4} e^{i2\pi f_s t} df_s$$

and are given by

$$\langle\mu\rangle = \frac{\int_{f_s=0}^{f_s=f_{max}} Re(b)\frac{f_s}{f_N} df_s}{\sqrt{P_{SOA,1}} - \sqrt{P_{SOA,0}}}$$

and

$$\langle\gamma\rangle = \frac{\int_{f_{s1}=-f_{max}}^{f_{s1}=f_{max}} \int_{f_{s2}=-f_{max}}^{f_{s2}=f_{max}} \left|\frac{f_{s1} - f_{s2}}{f_N}\right| \left(Re\left(b(f_{s1})\right)Re\left(b(f_{s2})\right) + Im\left(b(f_{s1})\right)Im\left(b(f_{s2})\right)\right) df_{s1}df_{s2}}{4\left(\sqrt{P_{SOA,1}} - \sqrt{P_{SOA,0}}\right)^2}$$

In the case where the signal PSD corresponds to a sinc function truncated for frequencies above $f_{max}$, $\langle\mu\rangle$ takes the simple form

$$\langle\mu\rangle = \frac{\int_0^{\pi\frac{f_{max}}{2f_N}} \frac{sin(x)}{x}\frac{2x}{\pi} dx}{\int_0^{\pi\frac{f_{max}}{2f_N}} \frac{sin(x)}{x} dx}$$

The case where an optical filter with a passband $2f_{OF}$ is interposed between the amplifier and the Rx is also treated, with the case $f_{OF} \leq f_F$ also resulting in a relatively simple description of the post electrical filtering level dependent noise

$$\sigma_{0/1} = \sqrt{2G^2Fhf_0 \cdot \left[P_{SOA,0/1}f_{OF} \pm \frac{1}{4}\left(P_{SOA,1} - P_{SOA,0}\right)\mu f_N - \frac{1}{8}\left(\sqrt{P_{SOA,1}} - \sqrt{P_{SOA,0}}\right)^2 \left(\mu + \frac{\gamma}{2}\right) f_N\right]}$$

Here $P_{SOA,0/1}$ are redefined as the 0- and 1-bit power levels taking ISI induced by the optical filter into account by referring it back to the input of the SOA. They are obtained by recording the 0- and 1- signal levels at the output of the SOA and dividing them by the SOA gain. Reduced extinction due to optical filtering in particular has to be taken into account in the numerical evaluation of the formulas. Pattern dependent effects are further discussed in the text.

In conclusion we have derived a set of equations describing the effect of optical and electrical filtering on level dependent ASE-signal beat noise in a non-return to zero ASK signal. Importantly, this effect results in a further reduction of signal quality that should be taken into account for accurate system modelling. The equations were validated by "brute force" numerical modeling of noisy data streams and comparison of the extracted and predicted noise levels after opto-electronic transduction and electrical filtering. Furthermore, the general trends of data dependent noise penalties were experimentally verified. The equations can be straightforwardly extended to other sources of noise.

## Acknowledgements

The authors would like to acknowledge funding for the project "Broadband Integrated and Green Photonic Interconnects for High-Performance Computing and Enterprise Systems" (BIG PIPES) funded under the seventh framework of the European Commission under contract no. 619591.